\documentclass[epj,twocolumn]{webofc}
\usepackage[varg]{txfonts}   
\usepackage{aas_macros} 
\usepackage{amssymb}    

\woctitle{CGS15}
\begin{document}
\title{The impact of global nuclear mass model uncertainties on $r$-process abundance predictions}
%
\author{M. Mumpower\inst{1}\fnsep\thanks{\email{matt.mumpower@nd.edu}} \and
        R. Surman\inst{1}\fnsep\thanks{\email{rsurman@nd.edu}} \and
        A. Aprahamian\inst{1}\fnsep\thanks{\email{aapraham@nd.edu}}
}
\institute{University of Notre Dame, Notre Dame, IN 46556, USA
}

\abstract{
Rapid neutron capture or `$r$-process' nucleosynthesis may be responsible for half the production of heavy elements above iron on the periodic table. 
Masses are one of the most important nuclear physics ingredients that go into calculations of $r$-process nucleosynthesis as they enter into the calculations of reaction rates, decay rates, branching ratios and Q-values. 
We explore the impact of uncertainties in three nuclear mass models on $r$-process abundances by performing global monte carlo simulations. 
We show that root-mean-square (rms) errors of current mass models are large so that current $r$-process predictions are insufficient in predicting features found in solar residuals and in $r$-process enhanced metal poor stars. 
We conclude that the reduction of global rms errors below $100$ keV will allow for more robust $r$-process predictions. 
}
\maketitle
\section{Introduction}\label{intro}
The cataclysmic event(s) responsible for the production of the heaviest neutron-rich elements found in nature is referred to as the rapid neutron capture process or $r$ process of nucleosynthesis. 
A complete description of the $r$ process still remains elusive with many open questions; chiefly among them is the unknown astrophysical location or site of the $r$ process. 
Proposed candidate sites include various scenarios that may occur in supernova and neutron star mergers, see Ref. \cite{Arnould2007} for a recent review. 

Strongly coupled to this longstanding problem is the pursuit of measurements on neutron-rich nuclei. 
Calculations of $r$-process nucleosynthesis require an appreciation of nuclear physics inputs for thousands of hard to measure, short-lived neutron-rich nuclei.
When measurements are not available theoretical extrapolations must be used which include predictions of nuclear masses (Q-values), $\beta$-decay rates, $\beta$-delayed neutron emission probabilities, neutron capture rates and photo-dissociation rates. 
Nuclear masses are conceivably the most important input since they contribute to all the other quantities. 
Nuclear masses also factor into the properties of the heaviest fissioning nuclei (e.g. barrier heights and daughter distributions) and further contribute to calculations of energy generation in astrophysical environments \cite{Schatz2013}. 

In a previous research campaign we performed investigations of individual nuclear properties. 
Individual neutron capture rates have been studied in the context of both weak \cite{AIP-NC} and main $r$-process components \cite{Mumpower2012c}. 
Studies of $\beta$-decay rates have been performed by us for the entire main $r$-process components \cite{AIP-Beta}. 
The impact of individual nuclear masses and binding energies was explored in Ref. \cite{Brett2012} and we showed the robustness of our conclusions under various astrophysical conditions in Refs. \cite{INPC2013,ICFN5}. 
We have since fused the efforts of these separate approaches by exploring how uncertainties in unmeasured nuclear masses propagate to all other quantities \cite{Mumpower2014a,Mumpower2014b}. 
By combining the abundance patterns of individual sensitivity studies we were able to approximate error bars on the $r$ process and concluded that root-mean-squared (rms) errors of global mass models should be around $100$ keV in order to minimize uncertainties in the predictions of $r$-process abundances. 

In this contribution we explore a new approach to assigning error bars on $r$-process abundance patterns from uncertain nuclear masses. 
Using a global monte carlo variation of all uncertain nuclear masses we create an ensemble of abundance patterns which are combined to produce the estimated error bars. 
This methodology reinforces the conclusions we drew from our studies of individual masses; that a reduction of global rms errors below $100$ keV will allow for accurate $r$-process predictions and differentiation between model predictions. 
An additional advantage of this new approach is that it will provide direct insight into correlations in nuclear masses. 

\section{Models}\label{model}
For this work we use a parameterized model from our previous investigations to study the impact of nuclear masses in the context of a `hot' $r$-process wind \cite{Mumpower2014a}. 
This trajectory has entropy of $200$ $k_B$, electron fraction of $Y_e=0.3$ and timescale of $80$ ms yielding the production of a main $r$-process component up to the third ($A=195$) peak. 
Hot $r$-process winds can achieve extreme conditions which produce a long equilibrium phase between neutron capture and photo-dissociation. 
As the supply of free neutrons is exhausted the $r$-process path, or set of most abundant isotopes moves back towards stability. 
It is during the decay back to stability that the equilibrium between neutron captures and photo-dissociations breaks down and individual rates begin to contribute to the predicted final abundances. 

Assuming detailed balance, nuclear masses enter into the calculation of photo-dissociation rates through one neutron separation energies:
\begin{equation}\label{eqn:photo}
\lambda_\gamma(Z,A) \propto T^{3/2} \exp\left[-{\frac{S_n(Z,A)}{kT}}\right] \langle \sigma v \rangle_{(Z,A-1)}
\end{equation}
where $S_n(Z,A)= M(Z,A-1)-M(Z,A)+M_n$ is the one neutron separation energy, $M(Z,A)$ is the mass of the nucleus with $Z$ protons and $A$ nucleons, $M(Z,A-1)$ is the mass of the nucleus with one less neutron, $M_n$ is the mass of the neutron, $T$ is the temperature, $\langle \sigma v \rangle_{(Z,A-1)}$ is the neutron capture rate of the neighboring nucleus and $k$ is Boltzmann's constant. 
A discussion of the photo-dissociation effect in the context of a hot wind $r$ process can be found in Ref. \cite{Surman2009}.

We calculate $r$-process abundances with a reduced reaction network used most recently in Refs. \cite{AIP-BE,AIP-Beta,AIP-NC}. 
To greatly speed up our network calculations we begin our simulations after seed nuclei have formed and charged particle reactions have frozen out, around $T=2$ GK. 
In addition we do not include fission recycling because our choice of trajectory is not sufficiently neutron-rich. 
Both of these approximations allow for increased statistics (they maximize the number of monte carlo steps) for our simulations in the context of a hot wind without altering our conclusions. 
We also note that a consequence of the reduced reaction network is that we only investigate the impact of masses above $A=20$. 

We explore three nuclear mass models with a range of physical assumptions: Finite Range Droplet Model (FRDM1995) \cite{FRDM1995}, Duflo Zuker (DZ) \cite{DZ} and Hartree-Fock-Bogoliubov version 17 (HFB-17) \cite{HFB17}. 
FRDM1995 is based on a macroscopic-microscopic approach which uses a finte-range liquid-droplet model and fold-Yukawa potential. The Strutinsky method is used for shell corrections. 
The Duflo Zuker model offers an alternative approach by starting from a shell-model monopole Hamiltonian. 
Using a pre-assumed shell structure terms in this model are calculated algebraically. 
The HFB models are entirely microscopic built on Skyrme (effective) interactions and realistic contact-pairing forces. 
We find these mass models have an rms error of $649$ keV, $405$ keV and $561$ keV respectively when compared to nuclei with $A > 20$ in the latest Atomic Mass Evaluation (2012 AME) \cite{AME2012}. 
To complete our nuclear network inputs these mass models are coupled with calculations of $\beta$-decay rates and neutron emission probabilities from the publicly available compilation \cite{Moller2003} and neutron capture rates from \cite{NONSMOKER1998}. 

We compare our final abundance patterns to elemental abundances from five metal poor stars: CS22892-052 \cite{CS22892}, CS31082-001 \cite{CS31082}, HD115444 \cite{HD115444}, HD221170 \cite{HD221170}, and BD+17\textdegree3248 \cite{BD17}, as in \cite{Mumpower2012b}. 
These metal poor halo stars are highly enriched in $r$-process elements and are believed to contain only a few events; which provides an excellent diagnostic for $r$-process models. 
We also compare our final abundance patterns to solar $r$-process residuals which are well constrained for heavy nuclei \cite{Arlandini1999}. 
The benefit of comparing $r$-process predictions to solar abundances is the use of isotopic data. 

\section{Monte Carlo}\label{mc}
\begin{figure}
\centerline{\includegraphics[width=9cm]{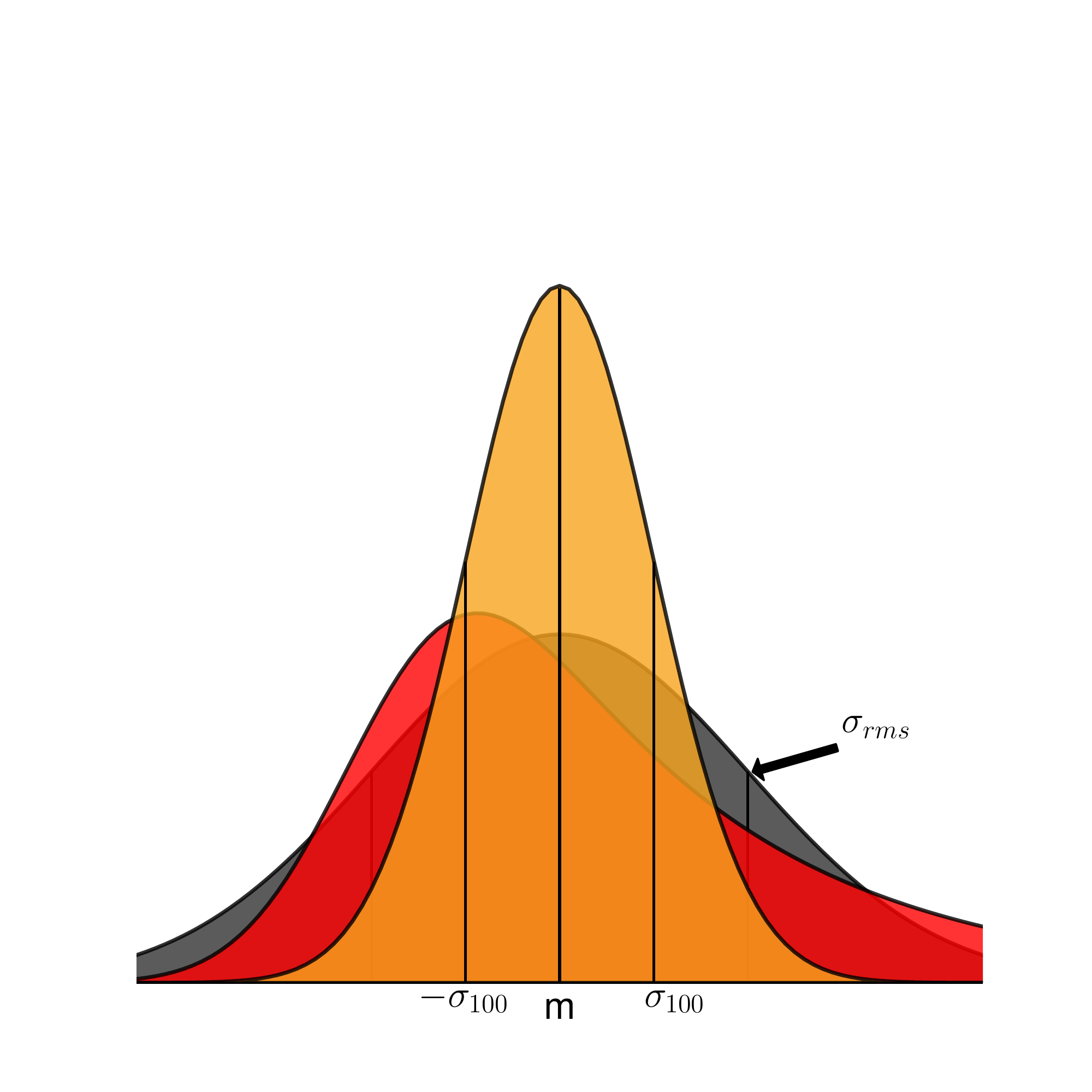}}
\caption{Distributions used to sample nuclear masses for our monte carlo studies. The gray curve represents a distribution centered on the theoretical mass value, $m$, with variance equal to a current mass model rms error, $\sigma_{rms}$. The the yellow curve represents a hypothetical improved model with a reduced rms error of $100$ keV, $\sigma_{100}$. The red curve is a skewed distribution that may be applicable to our monte carlo studies if measurements of neutron-rich masses show a systematic trend compared to theoretical extrapolations.}
\label{fig-1}
\end{figure}

To estimate error bars on $r$-process abundances we perform a monte carlo variation of uncertain nuclear masses. 
We first fix the aforementioned astrophysical hot wind $r$-process trajectory and the nuclear model, which sets the initial distribution of theoretical masses. 
For a given step, each nuclear mass that enters into the reaction network is varied using the probability distribution:
\begin{equation}\label{eqn:delta}
\Delta(Z,A) = \frac{1}{\sigma\sqrt{2\pi}} \exp\left[-\frac{(x-m)^2}{2\sigma^2}\right]
\end{equation}
where $m$ is the predicted mass from the theoretical model for nucleus with $Z$ protons and $A$ nucleons, $\sigma$ is the rms error when comparing the theoretical model to the 2012 AME and $x$ is a random variable. 
Thus, nuclear masses are sampled from a Gaussian distribution centered around the predicted theoretical value, $m$, of the given mass model with variance $\sigma_{rms}$ as shown in Fig. \ref{fig-1}. 
In this formulation mass variations above and below theoretical values are sampled with equal weight. 
Future mass measurements however may find general trends which cannot be adequately handled by this type of probability distribution. 
Instead a modified distribution may be needed such as the red curve in Fig. \ref{fig-1}. 
We note that mass variations are limited in our studies by excluding nuclei with experimentally measured values apart of the 2012 AME. 

Once this distribution of masses is generated we run our $r$-process simulation to record the final isotopic and elemental abundance patterns. 
This procedure is repeated thousands of times to gain sufficient statistics to create an ensemble of abundance patterns. 
To increase the number of steps in our monte carlo studies we make a final approximation by only recalculating one neutron separation energies which go into the calculation of photo-dissociation rates, see Equation \ref{eqn:photo}. 
This approximation has been shown to be valid for a hot wind $r$-process \cite{Mumpower2014a}, however if other astrophysical conditions are studied then the propagation of mass variations to all relevant quantities is needed \cite{Mumpower2014b}. 

\section{Results}\label{results}
\begin{figure*}
\centerline{\includegraphics[width=18cm,clip]{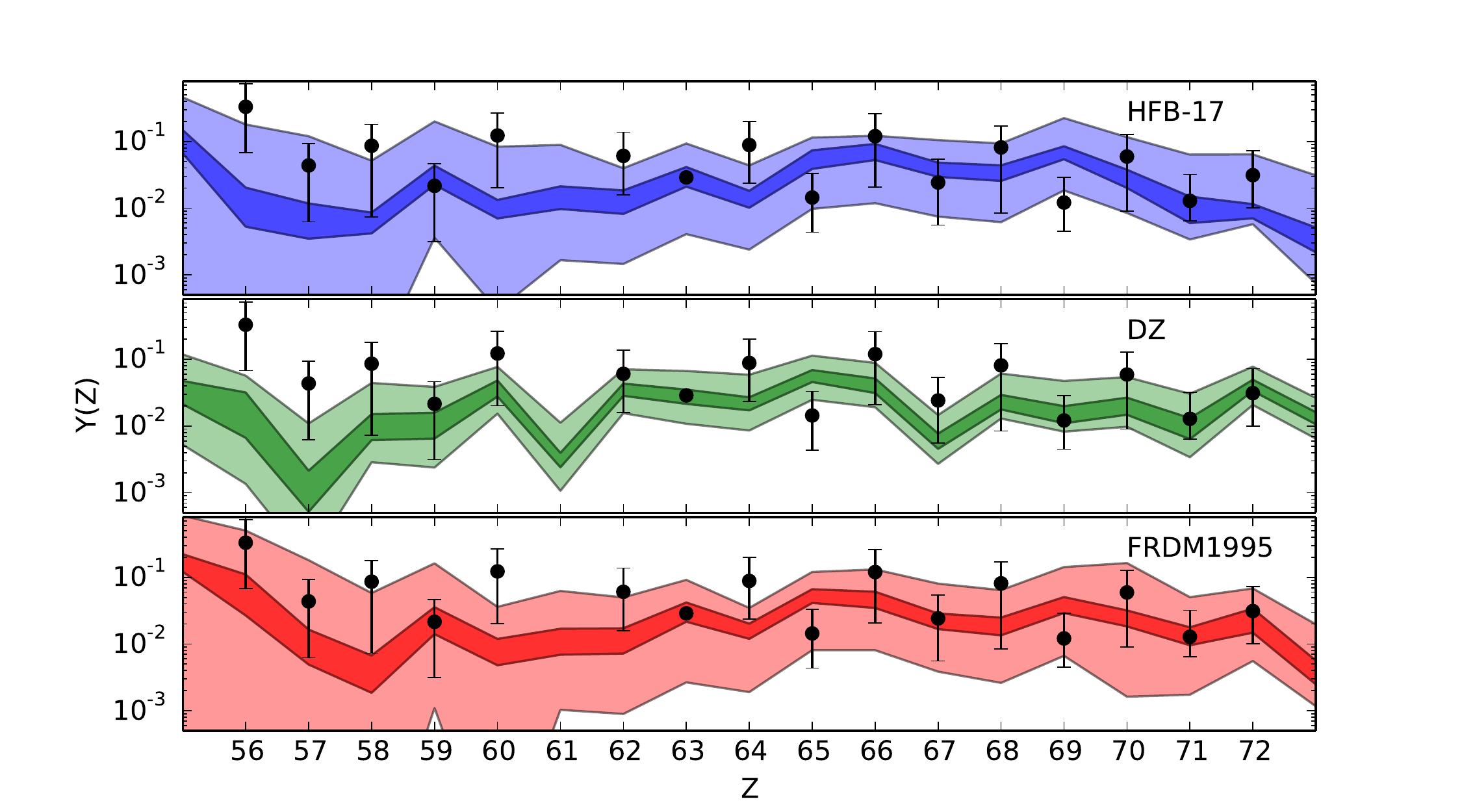}}
\caption{Variance in elemental abundance patterns for three nuclear models (HFB-17, DZ, and FRDM1995) compared to the average of five $r$-process enriched metal poor stars (dots). 
Darker shaded band represents the same monte carlo simulation with each mass model rms error fixed at $100$ keV.}
\label{fig-2}
\end{figure*}

We estimate uncertainty in our $r$-process predictions by computing the standard deviation (variance) of the isotopic and elemental abundance ensembles respectively. 
We further separate the ensemble data by the nuclear model used to calculate the final abundances. 

We highlight the results of our monte carlo studies for elemental abundance predictions, $Y(Z)$, in Fig. \ref{fig-2}. 
We scale each study to the value of Europium ($Z=63$) and also note the same scaling applies to the average of the five halo stars; hence there is no observational constraint for this data point (black dot). 
In these studies three nuclear mass models (HFB-17, DZ, and FRDM1995) were used and nuclear masses were varied globally using the prescription defined in the previous section. 

The lighter shaded color band represents the monte carlo study performed with the theoretical rms error of the given mass model. 
Despite vastly different physical assumptions, the abundance predictions for each mass model lie within the observational constraints. 
This suggests that with current mass model extrapolations we cannot confidently draw conclusions regarding $r$-process conditions. 
A second consequence of this result is the converse: we cannot use this astrophysical environment to differentiate between the three theoretical mass model extrapolations. 

The darker, bolder shaded color band in Fig. \ref{fig-2} represents an additional monte carlo study performed with a reduced rms model error of $100$ keV for each model. 
If mass model rms errors are hypothetically reduced then we begin to see differences in the elemental abundance predictions. 
For instance, predictions of Neodymium ($Z=60$) are only achieved with DZ while predictions of Lanthanum ($Z=57$) are only achieved with HFB-17 and FRDM1995. 
All mass models with reduced rms errors have abundance predictions that lie outside the observational constraints for Terbium ($Z=65$). 
In the context of these astrophysical conditions this suggests that either nuclear masses or rates in this region may suffer from systematic trends. 

\begin{figure*}
\centerline{\includegraphics[width=20cm,clip]{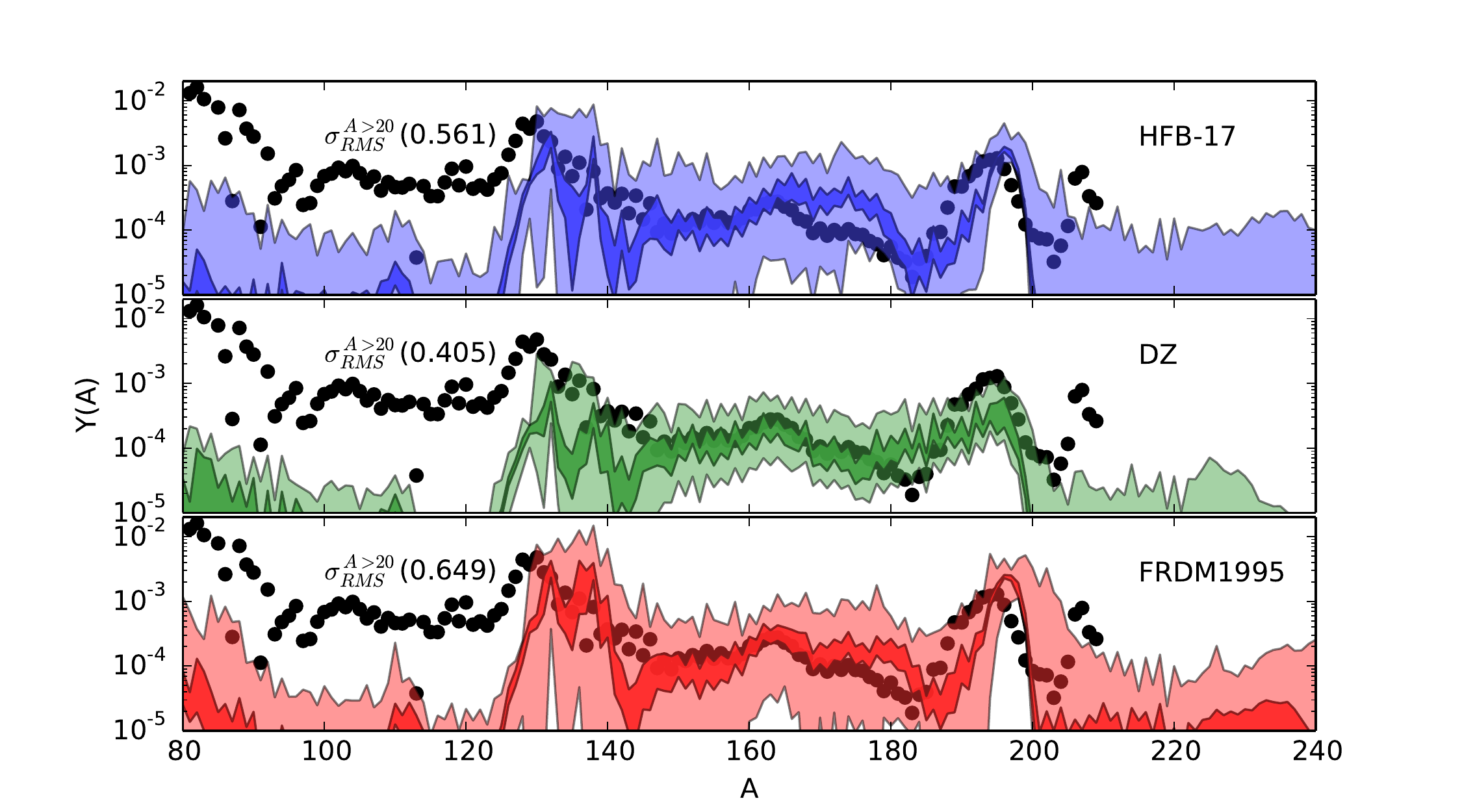}}
\caption{Variance in isotopic abundance patterns from three uncertain nuclear mass model predictions (HFB-17, DZ, and FRDM1995) compared to the solar $r$-process residuals (dots). 
Darker shaded band represents the same monte carlo simulation with each mass model rms error fixed at $100$ keV.}
\label{fig-3}
\end{figure*}

In Fig. \ref{fig-3} we show the results of our monte carlo studies for isotopic abundance predictions, $Y(A)$. 
We scale each study to best match the rare earth region ($A=150$ to $A=180$) of the solar pattern. 
As in Fig. \ref{fig-2}, the lighter shaded color band represents the monte carlo study performed with theoretical mass model rms errors and the darker shaded color band with a reduced model rms error of $100$ keV. 

The results of isotopic abundances echo what we found in the study of elemental abundances; that current theoretical mass model extrapolations lead to large error bars on our predictions of $Y(A)$. 
Only with a reduction of mass model errors, to $100$ keV, are we able to distinguish between the abundance predictions from different mass model extrapolations. 
For instance, with reduced mass model errors, we are able to discern the effects of late-time neutron capture which is present in both HFB-17 and FRDM1995 \cite{Mumpower2012b}. 

We find substantial variation for lighter nuclei in the second peak ($A\sim130$) region even with reduced mass model errors of $100$ keV due to the importance of the photo-dissociation effect near the $N=82$ and $Z=50$ closed shells which has been discussed previously \cite{Surman2009}. 
This result strongly suggests that nuclear masses in this region need to be known more accurately than $100$ keV. 
The weakest dependence on nuclear masses is seen in the third peak ($A\sim195$) region which provides additional support to the conclusions drawn in Ref. \cite{Mumpower2014b}. 

\section{Conclusions}\label{conclusions}
We have used a monte carlo approach to estimate the variability of $r$-process predictions from uncertain nuclear masses in the context of a hot wind $r$-process. 
In this new approach we address the impact of \textit{global} mass model uncertainties by randomly varying all nuclear masses that enter into our network calculation using a Gaussian distribution centered on a theoretical value with variance equal to the rms error of the mass model, as depicted in Fig. \ref{fig-1}. 
For each distribution of nuclear masses we simulate the $r$-process and record the final composition, creating an ensemble of abundances. 
The procedure developed here allows us to put error bars on abundance predictions by computing the variance of this ensemble. 

We have shown that current mass model extrapolations produce large variability in $r$-process predictions which can range orders of magnitude. 
The dramatic feature of this variability is that it come exclusively from mass variations of unmeasured nuclei, just beyond the reach of current radioactive beam facilities. 
The size of the variability is large enough that it is presently difficult to distinguish between mass model extrapolations for given astrophysical conditions. 

In fact, our approach provides a lower bound on the impact of uncertain nuclear masses on $r$-process abundance predictions since it does not account for the changes to other nuclear properties when a nuclear mass is varied. 
Additionally, it is well known that theoretical mass extrapolations vary by many MeV as neutron excess increases, see Fig. 1 and 2 in Ref. \cite{Mumpower2014b}. 
Thus, in the context of current model predictions at the neutron dripline, it may be that the choice of mass model rms error, $\sigma_{rms}$, for the variance of the Gaussian distribution is potentially too small. 
This is a point of concern for $r$-process trajectories where the path extends to the most neutron-rich nuclei, e.g. in neutron star mergers. 
For the hot wind studied here, the path stays closer to stability due to the long equilibrium phase, and so does not reach the dripline. 

To gauge the minimum mass model rms error needed for accurate abundance predictions we performed additional monte carlo studies. 
We found that a reduction of mass model uncertainties to $100$ keV suggests that variability in $r$-process predictions can be reduced to less than an order of magnitude in most regions. 
This result reinforces the conclusions we made when exploring the influence of individual nuclear masses \cite{Mumpower2014a,Mumpower2014b}. 
Furthermore, it provides extremely strong motivation for the continued effort of developing mass measurement campaigns at radioactive beam facilities as well as a renewed focus in improving global nuclear mass models. 
We plan to broaden this work in the future by exploring the effects of correlations, skewed mass distributions and expanding to additional astrophysical environments by propagating uncertainties to other nuclear physics quantities when nuclear masses are varied. 

\section{Acknowledgements}\label{acknowledgements}
We thank Sanjay Reddy, Gabriel Martinez-Pinedo, Peter M{\"o}ller and John Hardy for helpful discussions. 
This work was supported by the NSF through the Joint Institute for Nuclear Astrophysics grant numbers PHY0822648 and PHY1419765. 

\bibliographystyle{unsrt}
\bibliography{refs}

\begin{thebibliography}{25}

\bibitem{Arnould2007}
M.~Arnould, S.~Goriely, K.~Takahashi, Physics Reports \textbf{450}, 97  (2007)

\bibitem{Schatz2013}
H.~{Schatz}, International Journal of Mass Spectrometry \textbf{349}, 181
  (2013)

\bibitem{AIP-NC}
R.~{Surman}, M.~{Mumpower}, R.~{Sinclair}, K.L. {Jones}, W.R. {Hix}, G.C.
  {McLaughlin}, AIP Advances \textbf{4}, 041008 (2014)

\bibitem{Mumpower2012c}
M.~Mumpower, G.C. McLaughlin, R.~Surman, Phys.Rev. \textbf{C86}, 035803 (2012)

\bibitem{AIP-Beta}
M.~{Mumpower}, J.~{Cass}, G.~{Passucci}, R.~{Surman}, A.~{Aprahamian}, AIP
  Advances \textbf{4}, 041009 (2014)

\bibitem{Brett2012}
S.~{Brett}, I.~{Bentley}, N.~{Paul}, R.~{Surman}, A.~{Aprahamian}, European
  Physical Journal A \textbf{48}, 184 (2012)

\bibitem{INPC2013}
R.~{Surman}, M.~{Mumpower}, J.~{Cass}, I.~{Bentley}, A.~{Aprahamian}, G.C.
  {McLaughlin}, in \emph{European Physical Journal Web of Conferences} (2014),
  Vol.~66 of \emph{European Physical Journal Web of Conferences}, p. 7024

\bibitem{ICFN5}
R.~{Surman}, M.R. {Mumpower}, J.~{Cass}, A.~{Aprahamian}, \emph{{The
  Sensitivity of r-PROCESS Nucleosynthesis to the Properties of Neutron-Rich
  Nuclei}}, in \emph{ICFN5. World Scientific Publishing Co. ISBN
  9789814525435}, edited by J.H. {Hamilton}, V.~{Ramayya Akunuri} (2014), pp.
  538--545, \texttt{1309.0058}

\bibitem{Mumpower2014a}
M.~Mumpower, D.~Fang, R.~Surman, M.~Beard, A.~Aprahamian, submitted  (2014)

\bibitem{Mumpower2014b}
M.~Mumpower, R.~Surman, D.~Fang, M.~Beard, A.~Aprahamian, J. Phys. G accepted
  (2014)

\bibitem{Surman2009}
R.~{Surman}, J.~{Beun}, G.C. {McLaughlin}, W.R. {Hix}, \prc \textbf{79}, 045809
  (2009)

\bibitem{AIP-BE}
A.~{Aprahamian}, I.~{Bentley}, M.~{Mumpower}, R.~{Surman}, AIP Advances
  \textbf{4}, 041101 (2014)

\bibitem{FRDM1995}
P.~{M{\"o}ller}, J.R. {Nix}, W.D. {Myers}, W.J. {Swiatecki}, Atomic Data and
  Nuclear Data Tables \textbf{59}, 185 (1995)

\bibitem{DZ}
J.~{Duflo}, A.P. {Zuker}, \prc \textbf{52}, 23 (1995)

\bibitem{HFB17}
S.~{Goriely}, N.~{Chamel}, J.M. {Pearson}, Physical Review Letters
  \textbf{102}, 152503 (2009)

\bibitem{AME2012}
W.~{M.}, G.~{Audi}, W.~{A.~H.}, K.~{F.~G.}, M.~{MacCormick}, X.~{Xu},
  B.~{Pfeiffer}, Chinese Physics C \textbf{36}, 3 (2012)

\bibitem{Moller2003}
P.~{M{\"o}ller}, B.~{Pfeiffer}, K.L. {Kratz}, \prc \textbf{67}, 055802 (2003)

\bibitem{NONSMOKER1998}
T.~{Rauscher}, F.K. {Thielemann}, \emph{{Global statistical model calculations
  and the role of isospin}}, in \emph{Stellar Evolution, Stellar Explosions and
  Galactic Chemical Evolution}, edited by {A.~Mezzacappa} (1998), p. 519

\bibitem{CS22892}
C.~{Sneden}, A.~{McWilliam}, G.W. {Preston}, J.J. {Cowan}, D.L. {Burris}, B.J.
  {Armosky}, \apj \textbf{467}, 819 (1996)

\bibitem{CS31082}
V.~{Hill}, B.~{Plez}, R.~{Cayrel}, T.C. {Beers}, B.~{Nordstr{\"o}m},
  J.~{Andersen}, M.~{Spite}, F.~{Spite}, B.~{Barbuy}, P.~{Bonifacio} et~al.,
  \aap \textbf{387}, 560 (2002)

\bibitem{HD115444}
J.~{Westin}, C.~{Sneden}, B.~{Gustafsson}, J.J. {Cowan}, \apj \textbf{530}, 783
  (2000)

\bibitem{HD221170}
I.I. {Ivans}, J.~{Simmerer}, C.~{Sneden}, J.E. {Lawler}, J.J. {Cowan},
  R.~{Gallino}, S.~{Bisterzo}, \apj \textbf{645}, 613 (2006)

\bibitem{BD17}
J.J. {Cowan}, C.~{Sneden}, S.~{Burles}, I.I. {Ivans}, T.C. {Beers}, J.W.
  {Truran}, J.E. {Lawler}, F.~{Primas}, G.M. {Fuller}, B.~{Pfeiffer} et~al.,
  \apj \textbf{572}, 861 (2002)

\bibitem{Mumpower2012b}
M.~Mumpower, G.C. McLaughlin, R.~Surman, \apj \textbf{752}, 117 (2012)

\bibitem{Arlandini1999}
C.~{Arlandini}, F.~{K{\"a}ppeler}, K.~{Wisshak}, R.~{Gallino}, M.~{Lugaro},
  M.~{Busso}, O.~{Straniero}, \apj \textbf{525}, 886 (1999)

\end{thebibliography}

\end{document}